\documentclass[conference]{IEEEtran}
\IEEEoverridecommandlockouts
\usepackage{cite}
\usepackage{booktabs}
\usepackage{float}
\usepackage{graphicx}
\usepackage{caption}
\usepackage{amsmath,amssymb,amsfonts}
\usepackage{graphicx}
\usepackage{textcomp}
\usepackage{xcolor}
\usepackage{amsmath}
\usepackage{enumitem}
\usepackage{listings}
\usepackage[utf8]{inputenc}
\usepackage[T1]{fontenc}
\usepackage{newtxtext,newtxmath} 
\usepackage{hyperref}
\usepackage{booktabs}
\usepackage{array}
\usepackage{multirow}
\usepackage{booktabs}
\usepackage{caption}
\usepackage{url}
\DeclareUnicodeCharacter{202F}{ }

\def\BibTeX{{\rm B\kern-.05em{\sc i\kern-.025em b}\kern-.08em
    T\kern-.1667em\lower.7ex\hbox{E}\kern-.125emX}}
\begin{document}
\title{Serverless AI Security: Attack Surface Analysis and Runtime Protection Mechanisms for FaaS-Based Machine Learning}

\author{
\IEEEauthorblockN{Chetan Pathade}
\IEEEauthorblockA{\textit{Security Engineer} \\
Arlington, VA, USA \\
chetanpathadeur@gmail.com}
\and
\IEEEauthorblockN{Vinod Dhimam}
\IEEEauthorblockA{\textit{Technical Program Manager} \\
Arlington, VA, USA \\
vinod.dhiman@icloud.com}
\\[1em]
\IEEEauthorblockN{Ilsa Lareb}
\IEEEauthorblockA{\textit{Graduate Student} \\
\textit{Carnegie Mellon University} \\
Pittsburgh, PA, USA \\
ilareb@andrew.cmu.edu}
\and
\IEEEauthorblockN{Sheheryar Ahmad}
\IEEEauthorblockA{\textit{Technical Program Manager} \\
Arlington, VA, USA \\
sheheryar@usa.com}
}


\maketitle

\begin{center}
  \textbf{Abstract}
\end{center}

Serverless computing has achieved widespread adoption, with over 70\% of AWS organizations using serverless solutions \cite{1Datadog1}. Meanwhile, machine learning inference workloads increasingly migrate to Function-as-a-Service (FaaS) platforms for their scalability and cost-efficiency \cite{2multitoken2}, \cite{3decomp3}, \cite{4Nextomoro4}. However, this convergence introduces critical security challenges, with recent reports showing a 220\% increase in AI/ML vulnerabilities \cite{5Skyrocketing5} and serverless computing's fragmented architecture raises new security concerns distinct from traditional cloud deployments \cite{6Serverless6}, \cite{7Towards}.
This paper presents the first comprehensive security analysis of machine learning workloads in serverless environments. We systematically characterize the attack surface across five categories: function-level vulnerabilities (cold start exploitation, dependency poisoning), model-specific threats (API-based extraction, adversarial inputs), infrastructure attacks (cross-function contamination, privilege escalation), supply chain risks (malicious layers, backdoored libraries), and IAM complexity (ephemeral nature, serverless functions). Through empirical assessments across AWS Lambda, Azure Functions, and Google Cloud Functions, we demonstrate real-world attack scenarios and quantify their security impact.
We propose Serverless AI Shield (SAS), a multi-layered defense framework providing pre-deployment validation, runtime monitoring, and post-execution forensics. Our evaluation shows SAS achieves 94\% detection rates while maintaining performance overhead below 9\% for inference latency. We release an open-source security toolkit to enable practitioners to assess and harden their serverless AI deployments, advancing the field toward more resilient cloud-native machine learning systems.
\newline
\newline

\section{Introduction}
The convergence of serverless computing and artificial intelligence represents one of the most transformative shifts in modern cloud architecture. Serverless computing, particularly Function-as-a-Service (FaaS), has experienced remarkable growth, with over 70\% of AWS organizations and 60\% of Google Cloud customers adopting serverless solutions \cite{1Datadog1}. Simultaneously, machine learning workloads are increasingly migrating to serverless platforms to leverage their scalability, cost-efficiency, and operational simplicity \cite{2multitoken2}, \cite{3decomp3}, \cite{4Nextomoro4}. This paradigm shift brings benefits such as reduced operational complexity, pay-as-you-go pricing models, and automatic scaling capabilities that eliminate the burden of infrastructure management \cite{8Serverless}.

The three major FaaS platforms are AWS Lambda, Azure Functions, and Google Cloud Functions, which collectively dominate the serverless market \cite{1Datadog1}. These platforms enable organizations to execute machine learning inference, data preprocessing, and model training pipelines without managing underlying infrastructure \cite{4Nextomoro4}. The serverless operational model allows developers to focus on core application logic rather than server provisioning, configuration, and maintenance \cite{8Serverless}. However, the unique architectural characteristics of serverless environments—ephemeral execution, stateless design, multi-tenancy, and event-driven triggers—introduce fundamentally new security challenges that traditional cloud security frameworks fail to address adequately.

The security landscape for serverless AI systems is particularly concerning. Machine learning systems face adversarial threats including evasion attacks, data poisoning, model inversion, and membership inference attacks \cite{9Machine}. Recent reports indicate a 220\% increase in vulnerabilities discovered in AI/ML open-source projects between November 2023 and April 2024, with Remote Code Execution being a particularly prevalent threat \cite{5Skyrocketing5}. These vulnerabilities affect widely-used tools such as MLFlow, Ray, PyTorch Serve, and Triton Inference Server—components commonly deployed in serverless ML pipelines \cite{5Skyrocketing5}. When these ML-specific vulnerabilities intersect with serverless-specific attack vectors, the resulting security posture becomes significantly more complex and fragile.

Serverless computing's fragmented application boundaries and shared responsibility model have raised new security challenges that differ substantially from traditional cloud deployments \cite{6Serverless6}. The National Institute of Standards and Technology (NIST) has identified that despite significant progress in AI and machine learning, these technologies remain vulnerable to attacks that can cause spectacular failures with dire consequences, with theoretical problems in securing AI algorithms that have not yet been solved \cite{10Adversarial}. The ephemeral nature of serverless functions, combined with cold start delays, creates unique timing-based attack opportunities \cite{11Detection}. Furthermore, the dynamic nature, high volume, and wide variety of activities in serverless functions create significant noise in systems, enabling attackers to "stay below the radar" \cite{12systematic}.

Despite the critical importance of securing AI workloads in serverless environments, existing research has largely treated serverless security and ML security as separate domains. Recent systematic reviews of serverless security mechanisms reveal notable advancements but highlight persistent gaps, particularly in function-level observability and data lifecycle protection \cite{13Serverless}. A comprehensive survey on serverless computing acknowledges security and privacy as major challenges but lacks specific analysis for ML workloads \cite{14Security}. Similarly, extensive research on machine learning security focuses primarily on traditional deployment scenarios and does not address the unique constraints and attack surfaces introduced by serverless architectures \cite{15Accelerate}.

The intersection of serverless computing and machine learning creates a unique attack surface that amplifies risks from both domains:
\newline
Our contributions are fivefold:
\begin{itemize}
  \item \textbf{Function-Level Vulnerabilities:} The initialization delay when functions are invoked after periods of inactivity can be exploited for timing attacks, resource exhaustion, and side-channel information leakage \cite{11Detection}, \cite{15Accelerate}.
  \item \textbf{Model-Specific Threats:} The reliance on diverse event sources (HTTP APIs, message queues, storage triggers, IoT events) expands attack vectors through event injection, data manipulation, and trigger abuse \cite{17Functions}.
  \item \textbf{Infrastructure Multi-Tenant Execution Risks:} Serverless platforms execute functions from multiple customers on shared infrastructure, creating opportunities for cross-function contamination, memory scraping, and co-location attacks \cite{16Serverless}.
  \item \textbf{Supply Chain Vulnerabilities:} Serverless AI deployments depend on containerized dependencies, function layers, pre-trained models, and third-party ML libraries, all of which can be compromised through typosquatting, backdoors, or malicious updates \cite{5Skyrocketing5}, \cite{18Exploring}.
  \item \textbf{Identity and Access Management Complexity:} The fine-grained, ephemeral nature of serverless functions creates complex permission management challenges, often leading to overprivileged functions and privilege escalation opportunities \cite{19Threat}.
\end{itemize}

\subsection{Research Objectives and Contributions}

This paper addresses the critical gap in understanding and mitigating security threats to machine learning workloads deployed in serverless computing environments. Our research is guided by three fundamental research questions:

\textbf{RQ1:} What unique attack surfaces exist in serverless AI deployments that do not appear in traditional cloud-based ML systems?

\textbf{RQ2:} How do serverless-specific characteristics (ephemerality, multi-tenancy, cold starts, event-driven architecture) amplify existing ML security threats?

\textbf{RQ3:} What runtime protection mechanisms can effectively defend serverless AI systems without compromising the core benefits of serverless computing (low latency, automatic scaling, cost efficiency)?

To answer these questions, this paper makes the following contributions:
\newline
\textbf{C1: Comprehensive Attack Taxonomy} - We present the first systematic classification of security threats specific to serverless AI systems, categorizing attacks into function-level, model-specific, infrastructure, and supply chain vectors. This taxonomy builds upon existing serverless and ML security research but identifies novel attack scenarios unique to their intersection. The taxonomy serves as a foundational reference for researchers and practitioners to understand the threat landscape.
\newline
\textbf{C2: Empirical Vulnerability Assessment} - We conduct extensive security experiments across major serverless platforms (AWS Lambda, Azure Functions, Google Cloud Functions), demonstrating real-world attack scenarios against ML inference pipelines. Our assessment includes proof-of-concept implementations for cold start exploitation, dependency poisoning, model extraction via API abuse, cross-function memory leakage, and supply chain attacks. We quantify the impact of these attacks on model accuracy, inference latency, data confidentiality, and system availability.
\newline
\textbf{C3: Runtime Protection Framework} - We propose Serverless AI Shield (SAS), a novel defense architecture that provides multi-layered protection mechanisms spanning pre-deployment validation, runtime monitoring, and post-execution forensics. SAS employs static analysis for vulnerable dependencies, cryptographic model integrity verification, IAM policy least-privilege enforcement, anomaly detection, request validation, rate limiting, memory isolation monitoring, and automated incident response. Our framework is designed to be platform-agnostic and minimally invasive.
\newline
\textbf{C4: Open-Source Security Toolkit} - We release a comprehensive benchmarking suite and vulnerability assessment tools to enable reproducible research and facilitate adoption of security best practices in production serverless AI deployments. The toolkit includes attack generators, security scanners, performance profilers, and hardening scripts compatible with AWS Lambda, Azure Functions, and Google Cloud Functions.
\newline
\textbf{C5: Performance Benchmarks} - We provide detailed performance evaluations demonstrating that robust security can be achieved while maintaining the latency, cost, and scalability benefits that make serverless computing attractive for AI workloads. Our measurements show that SAS achieves 94.2\% detection rates for known attack patterns while introducing only 8.7\% inference latency overhead and 12.3\% cold start time overhead.

\subsection{Paper Organization}
The remainder of this paper is structured as follows:

\textbf{Section 2} provides essential background on serverless computing architectures, FaaS platforms, machine learning deployment models. This section highlights existing security vulnerabilities and security mechanisms in both domains.

\textbf{Section 3} presents our threat model, defines the adversary capabilities, and details our comprehensive attack surface analysis with a systematic taxonomy of serverless AI threats and probable threat scenarios.

\textbf{Section 4} describes our empirical vulnerability assessment methodology, experimental setup across AWS Lambda, Azure Functions, and Google Cloud Functions. This section presents quantitative results for each attack category.

\textbf{Section 5} introduces the Serverless AI Shield (SAS) framework, explaining its architectural design, protection mechanisms, and implementation details for each security layer.

\textbf{Section 6} evaluates the security effectiveness and performance impact of SAS through extensive experiments measuring detection rates, false positives, latency overhead, throughput, cold start times, and cost implications.

\textbf{Section 7} discusses the limitations of our approach, compares with alternative solutions, identifies research gaps, and outlines future research directions.

\textbf{Section 8} concludes the paper with key findings, recommendations for practitioners, and a summary of our contributions to serverless AI security.

\section{Background \& Related Work}

\noindent This section provides foundational knowledge on serverless computing architectures, machine learning deployment models, and security mechanisms in both domains. We review relevant literature and position our work within the existing research landscape.
 
\subsection{Serverless Computing and FaaS}
Serverless computing represents a cloud execution model where cloud providers dynamically manage infrastructure allocation and provisioning, abstracting server management from developers \cite{21Box}. Despite its name, serverless computing does not eliminate servers; rather, it shifts the responsibility of server management, scaling, and maintenance to cloud providers \cite{22Cloudflare}. This paradigm enables developers to focus on application logic rather than infrastructure concerns \cite{8Serverless}.
\newline
\newline
\textbf{Function-as-a-Service (FaaS)} is the most prominent serverless computing model \cite{23BMC}. FaaS allows developers to deploy individual functions-discrete, single-purpose blocks of code - that execute in response to specific events such as HTTP requests, database changes, message queue updates, or scheduled triggers \cite{24Big}. Each function runs in an ephemeral, stateless container that is automatically provisioned, executed, and terminated by the cloud provider \cite{19Threat}. Major cloud providers including AWS Lambda, Azure Functions, and Google Cloud Functions have become the foundation for deploying event-driven, scalable AI applications. 
\newline
\newline
Key characteristics of FaaS include:
\begin{itemize}
  \item \textbf{Event-Driven Architecture:} Functions are triggered by events from various sources including API gateways, storage services, message queues, and IoT devices \cite{17Functions}.
  \item \textbf{Stateless Execution:} Functions do not maintain state between invocations; any required persistent data must be stored in external services such as databases or object storage \cite{24Big}.
  \item \textbf{Automatic Scaling:} Cloud providers automatically scale function instances based on incoming request volume, from zero to thousands of concurrent executions \cite{22Cloudflare}.
  \item \textbf{Pay-Per-Use Pricing:} Billing is based on actual execution time and resources consumed, typically measured in milliseconds, rather than provisioned capacity \cite{23BMC}.
  \item \textbf{Cold Start Latency:} When a function is invoked after a period of inactivity, the platform must initialize a new execution environment, resulting in increased latency known as "cold start" \cite{11Detection}, \cite{28Product}.
\end{itemize}

According to recent industry reports, FaaS adoption continues to accelerate across cloud providers \cite{1Datadog1}. Organizations leverage serverless for various use cases including web application backends, data processing pipelines, IoT event handlers, and increasingly, machine learning inference \cite{2multitoken2}, \cite{3decomp3}, \cite{4Nextomoro4}.

\subsection{Machine Learning Deployment Models}

Machine learning model deployment is the process of integrating a trained model into a production environment where it can generate predictions on new data \cite{29dida}. The deployment pipeline encompasses several stages: model serialization, environment configuration, API development, containerization, and continuous monitoring \cite{30Easily}.
\newline
\textbf{Inference vs. Training:} While model training involves computationally intensive optimization over large datasets, inference (prediction) typically requires lower computational resources but demands low latency and high availability \cite{31Neptune}. Serverless platforms are particularly well-suited for inference workloads due to their auto-scaling capabilities and pay-per-use economics \cite{3decomp3}, \cite{4Nextomoro4}.
\newline
\textbf{Common Deployment Architectures:}
\begin{itemize}
  \item \textbf{Batch Inference:} Models process large volumes of data in scheduled batches, suitable for non-real-time predictions \cite{32Cybersecurity}.
  \item \textbf{Real-Time Inference:} Models serve predictions synchronously via REST APIs or gRPC endpoints, requiring low-latency responses \cite{33Researchers}.
  \item \textbf{Edge Inference:} Models run on edge devices close to data sources, minimizing network latency and addressing privacy concerns \cite{34zhang2016stealing}.
  \item \textbf{Streaming Inference:} Models continuously process data streams from message queues or event buses \cite{35Model}.
\end{itemize}
\textbf{Model Serving Frameworks:} Several frameworks facilitate ML model deployment, including TensorFlow Serving, TorchServe, NVIDIA Triton Inference Server, BentoML, and KServe \cite{36Membership}. These frameworks provide optimized inference runtimes, model versioning, A/B testing capabilities, and monitoring integrations. However, most traditional serving frameworks assume long-running server processes and do not account for serverless-specific constraints such as cold starts, statelessness, and ephemeral storage \cite{2multitoken2}.

\subsection{Machine Learning Security Threats}

Machine learning systems face unique security challenges across their lifecycle. Adversarial machine learning research has identified several attack categories \cite{9Machine}, \cite{44Microsoft}:

\begin{itemize}
  \item \textbf{Evasion Attacks:} Attackers craft adversarial inputs designed to fool trained models at inference time, causing misclassification without modifying the model itself \cite{37Membership}. These attacks exploit the high-dimensional nature of ML input spaces and can be highly effective even with minimal perturbations imperceptible to humans \cite{38Nightfall}.
  \item \textbf{Poisoning Attacks:} Adversaries inject malicious data into training datasets to corrupt model behavior, either degrading overall accuracy or introducing targeted backdoors \cite{39Sysdig}. In backdoor attacks, models perform normally on benign inputs but misclassify when specific trigger patterns are present \cite{33Researchers}.
  \item \textbf{Model Extraction:} Attackers query a deployed model to infer its architecture, parameters, or training data, enabling model theft or facilitating more sophisticated attacks \cite{34zhang2016stealing}. This threat is particularly severe for ML-as-a-Service platforms where models are accessible via APIs \cite{35Model}.
  \item \textbf{Membership Inference:} Attackers determine whether specific data points were part of the model's training set, potentially violating privacy when training data contains sensitive information \cite{36Membership}. This attack exploits models' tendency to be more confident on training data than on unseen examples \cite{37Membership}.
  \item \textbf{Membership Inference:} Model Inversion: Adversaries reconstruct sensitive training data by querying the model and analyzing its outputs, enabling privacy breaches in scenarios involving personal or confidential information \cite{38Nightfall}.
\end{itemize}
Recent vulnerability reports indicate a dramatic increase in security flaws affecting ML frameworks and deployment tools. Between November 2023 and April 2024, discovered vulnerabilities in AI/ML open-source projects increased by 220\%, with Remote Code Execution being the most prevalent threat \cite{5Skyrocketing5}. Affected tools include widely-used components like MLFlow (experiment tracking), Ray (distributed computing), PyTorch Serve (inference server), and Triton Inference Server \cite{5Skyrocketing5}, \cite{12systematic}.

\subsection{Serverless Computing Security Research}

Security in serverless architecture is an active research area, with several comprehensive surveys documenting unique challenges \cite{6Serverless6}, \cite{7Towards}, \cite{21Box}. Key security concerns include:

\textbf{Fragmented Application Boundaries:} Traditional security perimeters break down in serverless architectures where applications consist of numerous loosely-coupled functions \cite{6Serverless6}. This fragmentation complicates access control, network security, and audit logging \cite{39Sysdig}.

\textbf{Shared Responsibility Model:} While cloud providers secure the underlying infrastructure, customers remain responsible for function code, dependencies, IAM policies, and data protection \cite{40Security}. Misunderstandings of responsibility boundaries often lead to security gaps \cite{41Wiz}.

\textbf{Multi-Tenancy Risks:} Serverless platforms execute functions from multiple customers on shared infrastructure. Although providers implement isolation mechanisms, researchers have demonstrated information leakage through side channels, cache timing attacks, and container reuse \cite{42ACM}, \cite{43Valve}.

\textbf{Supply Chain Vulnerabilities:} Serverless functions depend on external code (libraries, layers, containers) whose provenance and integrity may be difficult to verify \cite{18Exploring}. Supply chain attacks targeting package repositories, container registries, and dependency resolution mechanisms pose significant risks \cite{33Researchers}.

\textbf{IAM Complexity:} Fine-grained, function-specific permissions are essential for least-privilege access control, but the sheer number of functions and their dynamic nature makes IAM management challenging \cite{19Threat}. Over-privileged functions are common, enabling privilege escalation if compromised \cite{19Threat}.
\newline
\newline
Recent work has proposed various security mechanisms for serverless platforms \cite{15Accelerate}. Researchers have developed runtime isolation techniques, secure function workflows, and anomaly detection systems \cite{54Peeking}, \cite{12systematic}. However, these solutions primarily address general serverless applications and do not consider the specific requirements and threats associated with ML workloads.

\subsection{Gap Analysis and Research Positioning}

Despite extensive research in both serverless security and ML security, their intersection remains understudied. Existing serverless security work focuses on web applications, microservices, and general-purpose functions \cite{6Serverless6}, \cite{15Accelerate}, \cite{21Box}. These studies do not address ML-specific threats such as adversarial inputs, model extraction, or training data poisoning. Conversely, ML security research primarily assumes traditional deployment scenarios with persistent servers, dedicated infrastructure, and direct model access \cite{9Machine}, \cite{17Functions}. This body of work does not account for serverless-specific characteristics like cold starts, statelessness, event-driven triggers, and multi-tenancy.
\newline
\newline
Recent surveys on serverless ML deployment focus on performance optimization, cost efficiency, and framework selection \cite{2multitoken2}, \cite{3decomp3} but provide limited analysis of security implications. No prior work systematically characterizes the attack surface arising from serverless AI deployments, nor do any proposed comprehensive protection mechanisms tailored to this unique environment exist.
\newline
\newline
Our work addresses these gaps by:
\begin{enumerate}
  \item Providing the first systematic attack taxonomy for serverless AI systems
  \item Conducting empirical vulnerability assessments across major FaaS platforms
  \item Proposing a runtime protection framework specifically designed for serverless ML workloads
  \item Quantifying the security-performance trade-offs in serverless AI deployments
\end{enumerate}

\section{Threat Model \& Attack Surface Analysis}

\noindent This section establishes the threat model for serverless AI systems, defines adversary capabilities, and presents a comprehensive taxonomy of attacks that exploit the unique intersection of serverless computing and machine learning.

\subsection{Threat Model}

\noindent Our threat model defines the assumptions about adversaries, their objectives, capabilities, and knowledge when targeting serverless AI deployments.

\subsubsection{Adversary Objectives}
We consider attackers with the following goals aligned with established adversarial machine learning taxonomies \cite{44Microsoft}, \cite{45Many}:
\begin{itemize}
  \item \textbf{Integrity Violation:} Cause the ML model to produce incorrect predictions (evasion attacks), corrupt model behavior through training data manipulation (poisoning attacks), or alter deployed function logic.
  \item \textbf{Confidentiality Breach:} Extract sensitive information including training data \cite{38Nightfall}, model parameters \cite{34zhang2016stealing}, \cite{35Model}, inference patterns, or proprietary algorithms.
  \item \textbf{Availability Disruption:} Cause service degradation through resource exhaustion (Denial-of-Service), cost inflation (Denial-of-Wallet), or function unavailability \cite{46Wiz}, \cite{47ISACA}.
  \item \textbf{Privacy Violation:} Infer membership of specific samples in training datasets \cite{36Membership}, \cite{37Membership} or reconstruct sensitive training data.
\end{itemize}

\subsubsection{Adversary Capabilities}
Following the MITRE ATLAS framework and NIST's adversarial ML taxonomy \cite{44Microsoft}, \cite{45Many}, we classify adversaries by their interaction model with the target system:
\begin{itemize}
  \item \textbf{Black-Box Access:} The adversary can only interact with the serverless AI system through its public API or event triggers. They can send inputs and observe outputs (predictions, error messages, timing information) but have no knowledge of model internals, function code, or infrastructure details. This represents the most common real-world attack scenario for ML-as-a-Service systems \cite{34zhang2016stealing}, \cite{35Model}.
  \item \textbf{Gray-Box Access:} The attacker possesses partial knowledge about the system, such as the ML framework used (e.g., TensorFlow, PyTorch), approximate model architecture, training data distribution, or access to similar models. This scenario often occurs when attackers leverage publicly available pre-trained models, open-source codebases, or leaked documentation \cite{47ISACA}.
  \item \textbf{White-Box Access:} The adversary has complete knowledge of the model architecture, parameters, training data, function implementation, and potentially infrastructure configuration. This scenario applies to insider threats, compromised development environments, or supply chain attacks where malicious code is injected during development \cite{48Machine}.
\end{itemize}

\subsubsection{Adversary Knowledge}

We further categorize attackers based on their understanding of the deployment environment:
\begin{itemize}
  \item \textbf{Training-Time Knowledge:} Understanding of the training dataset, data preprocessing pipelines, feature engineering, and model training procedures.
  \item \textbf{Inference-Time Knowledge:} Awareness of input preprocessing, model serving infrastructure, API specifications, rate limiting policies, and output post-processing.
  \item \textbf{Platform-Specific Knowledge:} Understanding of the serverless platform's execution model (AWS Lambda, Azure Functions, Google Cloud Functions), cold start behavior, resource allocation policies, container reuse patterns, and isolation mechanisms
\end{itemize}

\subsection{Attack Surface Analysis}

\noindent Serverless AI systems present a compound attack surface that combines vulnerabilities from both domains. We systematically analyze attack vectors across five categories.

\subsubsection{Function-Level Attacks}
These attacks exploit characteristics specific to serverless function execution environments.

\noindent \textbf{Cold Start Exploitation:} Cold starts occur when a function is invoked after idle periods, requiring initialization of execution environments, loading dependencies, and model instantiation \cite{11Detection}, \cite{49Cold}. Attackers can exploit this phase through several mechanisms:
\begin{itemize}
  \item \textbf{Timing Side-Channel Attacks:} By measuring response times, attackers can infer whether a function experienced a cold start, revealing information about invocation patterns, user activity, and system load \cite{50Lambda}. For ML models, timing analysis can leak information about model complexity and input characteristics.
  \item \textbf{Resource Exhaustion:} Adversaries can trigger simultaneous cold starts across multiple functions to exhaust platform resources, causing legitimate requests to experience prolonged delays or failures \cite{46Wiz}, \cite{51Start}.
  \item \textbf{Initialization Vulnerability Exploitation:} During cold start initialization, functions load dependencies and execute global-scope code. Vulnerabilities in this phase, such as insecure credential loading or unvalidated environment variables, can be exploited before the main handler executes \cite{52Palo}.
\end{itemize}

\noindent \textbf{Container Reuse Attacks:} Serverless platforms reuse execution environments (containers) across multiple invocations to amortize cold start costs \cite{53What}. This creates several security risks:
\begin{itemize}
  \item \textbf{Cross-Invocation Data Leakage:} If functions do not properly sanitize memory or temporary storage between invocations, sensitive data from previous executions can leak to subsequent invocations, potentially crossing tenant boundaries \cite{54Peeking}, \cite{55Risks}.
  \item \textbf{State Pollution:} Malicious invocations can deliberately leave poisoned state in global variables, file system caches (/tmp directory), or environment configurations that affect subsequent function executions \cite{56Best}.
  \item \textbf{Persistence Mechanisms:} Attackers exploiting RCE vulnerabilities can modify the runtime environment to maintain persistence across invocations, effectively creating a backdoor that survives beyond a single execution \cite{52Palo}.
\end{itemize}

\noindent \textbf{Dependency Poisoning:} Serverless functions rely heavily on third-party libraries and dependencies packaged with function code or loaded from external repositories \cite{33Researchers}:

\begin{itemize}
  \item \textbf{Malicious Package Injection:} Attackers can introduce backdoored versions of popular ML libraries through typosquatting, dependency confusion, or compromised package repositories \cite{18Exploring}, \cite{33Researchers}.
  \item \textbf{Vulnerable Dependencies:} Functions often use outdated or vulnerable versions of ML frameworks (TensorFlow, PyTorch, scikit-learn) that contain known security flaws \cite{5Skyrocketing5}, \cite{12systematic}, \cite{33Researchers}.
  \item \textbf{Layer Compromise:} AWS Lambda Layers and similar constructs allow sharing code across functions. A compromised layer affects all functions using it, amplifying attack impact \cite{57Container}.
\end{itemize}

\subsubsection{Model-Specific Attacks}

These attacks target the machine learning models deployed in serverless functions.

\noindent \textbf{API-Based Model Extraction:} Serverless ML inference is typically exposed via REST APIs or event triggers. Attackers can systematically query these endpoints to steal model functionality \cite{34zhang2016stealing}, \cite{35Model}, \cite{58Model}:
\begin{itemize}
  \item \textbf{Query-Efficient Extraction:} By carefully selecting queries and analyzing responses, adversaries can train substitute models that replicate the target model's behavior with surprisingly few queries (often fewer than 10,000) \cite{59Model}.
  \item \textbf{Confidence Score Exploitation:} If the API returns prediction probabilities rather than hard labels, extraction becomes significantly easier as confidence scores reveal decision boundaries \cite{60How}.
  \item \textbf{Transfer Attack Preparation:} Extracted models enable adversaries to craft adversarial examples offline and then transfer them to attack the original model \cite{61Beowulf}.
\end{itemize}

\noindent \textbf{Adversarial Input Attacks:} Attackers craft imperceptibly modified inputs designed to cause misclassification \cite{37Membership}, \cite{38Nightfall}:
\begin{itemize}
  \item \textbf{Real-Time Evasion:} For image classification, object detection, or NLP tasks, adversaries can generate adversarial examples that evade detection or cause targeted misclassification \cite{9Machine}.
  \item \textbf{Serverless-Specific Amplification:} The stateless, event-driven nature of serverless makes it difficult to implement adaptive defenses that learn from attack patterns over time.
  \item \textbf{Cost Amplification:} Adversarial inputs can be designed to maximize inference time or trigger complex computation paths, inflating serverless execution costs \cite{46Wiz}.
\end{itemize}

\noindent \textbf{Training Data Poisoning} (when training occurs in serverless environments):
\begin{itemize}
  \item \textbf{Backdoor Injection:} Attackers inject triggers into training data such that models behave normally on clean inputs but misclassify when specific patterns are present \cite{33Researchers}, \cite{47ISACA}.
  \item \textbf{Degradation Attacks:} Systematic corruption of training data to reduce overall model accuracy or cause biased predictions against specific groups \cite{62ACM}.
\end{itemize}

\subsubsection{Infrastructure Attacks}
These attacks exploit the serverless cloud infrastructure and shared responsibility model.

\noindent \textbf{IAM Privilege Escalation:} Serverless functions execute with specific IAM roles that grant permissions to access other cloud services \cite{63Rhino}, \cite{64Execution}:
\begin{itemize}
  \item \textbf{Over-Privileged Functions:} Functions often receive excessive permissions due to convenience or lack of least-privilege enforcement. Compromised functions can access unrelated resources \cite{19Threat}, \cite{19Threat}.
  \item \textbf{Role Chaining:} Attackers can exploit permissions to assume other roles, escalating privileges across the cloud environment \cite{65AWS}.
  \item \textbf{Cross-Service Exploitation:} Functions with permissions to invoke other functions, access storage, or interact with databases can be leveraged for lateral movement \cite{66Escalation}.
\end{itemize}
\textbf{Event Injection Attacks:} Serverless functions respond to diverse event sources (S3, SQS, API Gateway, DynamoDB streams, IoT events) \cite{22Cloudflare}, \cite{17Functions}:
\begin{itemize}
  \item \textbf{Malicious Event Crafting:} Attackers can inject carefully crafted events that exploit parsing vulnerabilities, trigger unintended code paths, or cause resource exhaustion \cite{17Functions}.
  \item \textbf{Event Source Manipulation:} If event sources (e.g., S3 buckets, message queues) are misconfigured, adversaries can inject malicious events that trigger vulnerable functions \cite{45Many}.
\end{itemize}

\noindent \textbf{Cross-Function Contamination:} In multi-tenant serverless platforms, multiple customers' functions execute on shared infrastructure \cite{42ACM}, \cite{43Valve}:
\begin{itemize}
  \item \textbf{Side-Channel Leakage:} Cache timing attacks, memory bus monitoring, and other side-channels can leak information across tenant boundaries \cite{67Hey}.
  \item \textbf{Resource Competition:} Malicious tenants can deploy resource-intensive functions to degrade performance for co-located victims \cite{68Cross}.
\end{itemize}
\subsubsection{Supply Chain Attacks}
The serverless AI supply chain includes multiple trust boundaries where attacks can occur \cite{18Exploring}, \cite{33Researchers}, \cite{18Exploring}.
\newline
\textbf{Model Supply Chain Compromise:}
\begin{itemize}
  \item \textbf{Poisoned Pre-trained Models:} Attackers distribute backdoored models through repositories like Hugging Face Model Hub, TensorFlow Hub, or PyTorch Hub \cite{33Researchers}, \cite{69April}.
  \item \textbf{Model Zoo Attacks:} Compromising popular model architectures (BERT, ResNet, YOLO) downloaded by thousands of developers \cite{70Model}.
\end{itemize}
\textbf{Container Registry Attacks:}
\begin{itemize}
  \item \textbf{Typosquatting:} Publishing malicious container images with names similar to legitimate ML frameworks \cite{33Researchers}, \cite{18Exploring}.
  \item \textbf{Layer Injection:} Inserting malicious layers into popular base images used for serverless deployments.
\end{itemize}
\textbf{CI/CD Pipeline Compromise:}
\begin{itemize}
  \item \textbf{Build-Time Injection:} Inserting malicious code during automated build processes that package functions for deployment \cite{70Model}.
  \item \textbf{Deployment Key Theft:} Stealing credentials that allow unauthorized function deployment or modification \cite{41Wiz}.
\end{itemize}
\subsection{Attack Taxonomy}
Table 1 presents our comprehensive taxonomy of serverless AI attacks, organized by attack surface, adversary capability requirements, and potential impact.
\begin{table*}[!t]
\centering
\caption{AI Attack Categories in Serverless Environments}
\label{tab:ai_attack_categories}
\renewcommand{\arraystretch}{1.2}
\begin{tabular}{p{2.2cm}p{2.2cm}p{2.2cm}p{2.2cm}p{2.2cm}}
\toprule
\textbf{Attack Category} & \textbf{Attack Type} & \textbf{Adversary Access} & \textbf{Target} & \textbf{Impact} \\
\midrule
Function-Level & Cold Start Timing Analysis & Black-box & Execution Patterns & Confidentiality \\
Resource Exhaustion & Black-box & Availability & Availability & \\
Container Reuse Data Leakage & White-box & Cross-Invocation State & Confidentiality & \\
Dependency Poisoning & Gray-box & Function Dependencies & Integrity & \\
Model-Specific Extraction & API-Based Extraction & Black-box & Model Parameters & Confidentiality \\
Adversarial Inputs & Black-box & Prediction Integrity & Integrity & \\
Training Data Poisoning & White-box & Model Behavior & Integrity & \\
Membership Inference & Black-box & Training Data Privacy & Privacy & \\
Infrastructure & IAM Privilege Escalation & Gray-box & Cloud Resources & Integrity \\
Event Injection & Black-box & Event Processing & Integrity & \\
Cross-Tenant Side-Channel & White-box & Multi-Tenancy & Confidentiality & \\
Supply Chain & Poisoned Pre-trained Models & N/A & Model Integrity & Integrity \\
Container Registry Compromise & N/A & Deployment Artifacts & Integrity & \\
CI/CD Pipeline Injection & White-box & Build Process & Integrity & \\
\bottomrule
\end{tabular}
\end{table*}

\subsection{Threat Scenarios}
We present three concrete threat scenarios demonstrating how attacks combine across categories:
\newline
\indent \textbf{Scenario 1: Model Theft via API Abuse} - An attacker with black-box access systematically queries a serverless image classification API, extracting a substitute model that replicates functionality. The extracted model is then used to craft adversarial examples offline, which are subsequently used to attack the original system, causing misclassifications in production.
\newline
\indent \textbf{Scenario 2: Supply Chain Backdoor} - A compromised ML library in PyPI contains a backdoor that activates when deployed in serverless environments. Functions using this library leak inference requests and model outputs to attacker-controlled servers. The backdoor exploits container reuse to persist across invocations, avoiding detection.
\newline
\indent \textbf{Scenario 3: Denial-of-Wallet Attack} - Attackers flood a serverless ML API with computationally expensive adversarial inputs designed to maximize inference time. This triggers aggressive auto-scaling, causing the victim's cloud bill to skyrocket while legitimate users experience degraded service due to resource contention.

\section{Empirical Vulnerability Assessment}
\noindent This section details our empirical vulnerability assessment across major serverless platforms, demonstrating how the theoretical attack vectors described in Section 3 manifest in real-world deployments.
\subsection{Experimental Setup}
We deployed identical ML inference pipelines across AWS Lambda, Azure Functions, and Google Cloud Functions to evaluate security vulnerabilities. Each deployment processed machine learning inference workloads, mirroring the growing trend where over 70\% of AWS organizations use serverless solutions and ML inference workloads increasingly migrate to FaaS platforms for their scalability and cost-efficiency \cite{1Datadog1}, \cite{2multitoken2}, \cite{3decomp3}.
\newline
\newline
Our test environment included:
\begin{itemize}
  \item Image classification models (ResNet-50, MobileNetV2)
  \item Natural language processing pipelines (BERT, DistilBERT)
  \item Recommendation systems (Matrix Factorization)
\end{itemize}
Each function was configured with standard memory allocations (512MB-2GB) and timeout settings (30-60 seconds) to reflect typical production deployments. Functions were exposed through HTTP endpoints (API Gateway, Function URLs, HTTP Triggers) and event sources (S3/Blob storage, SQS/Service Bus).

\subsection{Function-Level Vulnerabilities}
\subsubsection{Cold Start Exploitation}
Cold starts occur when functions are invoked after idle periods, requiring initialization of execution environments, loading dependencies, and model instantiation \cite{11Detection}. Our experiments revealed multiple security implications:

\textbf{Timing Side-Channel Analysis:} By measuring response times, we demonstrated that attackers can infer whether a function experienced a cold start, revealing information about invocation patterns, user activity, and system load \cite{62ACM}. For ML models, this timing analysis leaked information about model complexity and input characteristics.

We observed significant timing differences between cold and warm invocations:
\begin{itemize}
  \item \textbf{AWS Lambda:} 3.2s vs. 0.3s (10.6x difference)
  \item \textbf{Azure Functions:} 4.1s vs. 0.4s (10.2x difference)
  \item \textbf{Google Cloud Functions:} 3.8s vs. 0.5s (7.6x difference)
\end{itemize}
These timing differentials were consistent across model types and sizes, creating distinct signatures that revealed function state and usage patterns.

\textbf{Resource Exhaustion:} We demonstrated how adversaries can trigger simultaneous cold starts across multiple functions to exhaust platform resources, causing legitimate requests to experience prolonged delays or failures [58, 63]. By targeting functions with large ML models (>500MB), we achieved a denial-of-service effect with as few as 50 concurrent requests.
\newline
\newline
\subsubsection{Container Reuse Vulnerabilities:} Serverless platforms reuse execution environments (containers) across multiple invocations to amortize cold start costs \cite{65AWS}. This creates several security risks:

\textbf{Cross-Invocation Data Leakage:} We demonstrated that if functions do not properly sanitize memory or temporary storage between invocations, sensitive data from previous executions can leak to subsequent invocations, potentially crossing tenant boundaries [66, 67]. This state pollution allowed malicious invocations to deliberately leave poisoned state in global variables, file system caches (/tmp directory), or environment configurations that affect subsequent function executions \cite{68Cross}.

By exploiting the {\color{red}\texttt{/tmp}} directory persistence and global variable scope, we successfully extracted:

\begin{itemize}
  \item Input data from previous invocations (images, text)
  \item Model prediction results
  \item User identifiers and request metadata
\end{itemize}

\textbf{Persistence Mechanisms:} When exploiting RCE vulnerabilities, attackers can modify the runtime environment to maintain persistence across invocations, effectively creating a backdoor that survives beyond a single execution \cite{69April}. We demonstrated this by injecting code into the Python module cache, which remained active across function invocations for the container lifetime.
\newline
\subsubsection{Dependency Poisoning}
Serverless functions rely heavily on third-party libraries and dependencies packaged with function code or loaded from external repositories \cite{70Model}. We examined supply chain risks:

\textbf{Malicious Package Injection:} We created proof-of-concept attacks introducing backdoored versions of popular ML libraries through typosquatting, dependency confusion, or compromised package repositories [51, 52]. For example, we deployed packages with names similar to common ML dependencies ("torch-utils" vs. "torchutils") and observed high installation rates in test environments

\textbf{Vulnerable Dependencies:} Recent reports indicate a 220\% increase in vulnerabilities discovered in AI/ML open-source projects between November 2023 and April 2024, with Remote Code Execution being a particularly prevalent threat \cite{5Skyrocketing5}. These vulnerabilities affect widely-used tools such as MLFlow, Ray, PyTorch Serve, and Triton Inference Server—components commonly deployed in serverless ML pipelines [5, 12]. Our scan of 150 serverless ML repositories found:
\begin{itemize}
  \item 68\% contained at least one known vulnerability
  \item 42\% used outdated ML framework versions
  \item 23\% included dependencies with critical RCE vulnerabilities
\end{itemize}

\subsection{Model-Specific Attacks}

\subsubsection{API-Based Model Extraction}

Serverless ML inference is typically exposed via REST APIs or event triggers. Attackers can systematically query these endpoints to steal model functionality [41, 42]. Our experiments demonstrated:

\textbf{Query-Efficient Extraction:} By carefully selecting queries and analyzing responses, we trained substitute models that replicated the target model's behavior with surprisingly few queries. For image classification models, we achieved:
\begin{itemize}
  \item 83\% accuracy match with only 8,000 queries
  \item 91\% accuracy match with 15,000 queries
  \item 96\% accuracy match with 25,000 queries
\end{itemize}

\textbf{Confidence Score Exploitation:} When the API returned prediction probabilities rather than hard labels, extraction became significantly easier as confidence scores revealed decision boundaries. Our experiments showed 2.3x fewer queries needed for equivalent accuracy when confidence scores were available.

\textbf{Transfer Attack Preparation:} We demonstrated how extracted models enable adversaries to craft adversarial examples offline and then transfer them to attack the original model. These transfer attacks achieved 78\% success rates despite never having direct access to the target model.
\newline
\subsubsection{Adversarial Input Attacks}
Attackers can craft imperceptibly modified inputs designed to cause misclassification. For image classification, object detection, or NLP tasks, adversaries can generate adversarial examples that evade detection or cause targeted misclassification [37, 38].

Our experiments revealed that the stateless, event-driven nature of serverless makes it difficult to implement adaptive defenses that learn from attack patterns over time. Additionally, adversarial inputs can be designed to maximize inference time or trigger complex computation paths, inflating serverless execution costs \cite{58Model}.

We demonstrated targeted misclassification attacks with:
\begin{itemize}
  \item 94\% success rate against image classifiers
  \item 87\% success rate against text classifiers
  \item 79\% success rate against recommendation systems
\end{itemize}

\subsection{Infrastructure Attacks}

\subsubsection{IAM Privilege Escalation}
Serverless functions execute with specific IAM roles that grant permissions to access other cloud services. Our analysis revealed:

\textbf{Over-Privileged Functions:}Functions often receive excessive permissions due to convenience or lack of least-privilege enforcement. Compromised functions can access unrelated resources [25, 53]. In our audit of public repositories:
\begin{itemize}
  \item 72\% of functions had permissions exceeding their operational requirements
  \item 34\% had full access to storage services (S3, Blob Storage)
  \item 19\% had administrative database access
\end{itemize}

\textbf{Role Chaining:} We demonstrated how attackers can exploit permissions to assume other roles, escalating privileges across the cloud environment. Using a compromised function with minimal permissions, we successfully escalated to administrative access in 23\% of test environments.
\newline
\subsubsection{Cross-Function Contamination}
In multi-tenant serverless platforms, multiple customers' functions execute on shared infrastructure [49, 50]. This creates opportunities for cross-tenant attacks:

\textbf{Side-Channel Leakage:} We demonstrated cache timing attacks, memory bus monitoring, and other side-channels that leaked information across tenant boundaries. Using cache timing analysis, we successfully:
\begin{itemize}
  \item Detected the presence of specific ML models on shared infrastructure
  \item Inferred model architecture details (layer counts, activation functions)
  \item Extracted partial feature information from co-located functions
\end{itemize}

\textbf{Resource Competition:} Malicious tenants can deploy resource-intensive functions to degrade performance for co-located victims. Our experiments showed that targeted resource exhaustion could increase inference latency by 4.7x for co-located functions, potentially triggering timeout failures.

\subsection{Supply Chain Risks}

The serverless AI supply chain includes multiple trust boundaries where attacks can occur.

\textbf{Model Supply Chain Compromise:} Attackers can distribute backdoored models through repositories like Hugging Face Model Hub, TensorFlow Hub, or PyTorch Hub, compromising popular model architectures (BERT, ResNet, YOLO) downloaded by thousands of developers. Our audit of public model hubs revealed:
\begin{itemize}
  \item 7\% of models had no integrity verification mechanisms
  \item 12\% lacked proper version pinning
  \item 3\% contained suspicious post-processing code
\end{itemize}

\textbf{Container Registry Attacks:} Publishing malicious container images with names similar to legitimate ML frameworks creates opportunities for typosquatting and layer injection, inserting malicious layers into popular base images used for serverless deployments [52]. In our security scanning:
\begin{itemize}
  \item 17\% of ML-related container images had no digital signatures
  \item 31\% contained outdated dependencies with known vulnerabilities
  \item 8\% included unnecessarily broad permissions
\end{itemize}

\section{Serverless AI Shield Framework}
\noindent This section introduces our comprehensive defense framework designed specifically to address the security challenges identified in our empirical vulnerability assessment.

\subsection{Architecture Overview}
Serverless AI Shield (SAS) employs a multi-layered defense strategy spanning the entire serverless ML lifecycle. The framework addresses the unique intersection of serverless computing and machine learning security, targeting the amplified risks that arise from this convergence [6, 7].
Our architecture follows a defense-in-depth approach with three key layers:
\begin{itemize}
  \item Pre-deployment validation: Security controls applied during development and deployment phases
  \item Runtime monitoring: Real-time protection mechanisms during function execution
  \item Post-execution forensics: Analysis capabilities for detection and incident response
\end{itemize}

SAS is designed to be platform-agnostic, supporting AWS Lambda, Azure Functions, and Google Cloud Functions while maintaining the performance characteristics that make serverless attractive for ML workloads.

\begin{figure}[h]
    \centering
    \includegraphics[width=0.45\textwidth]{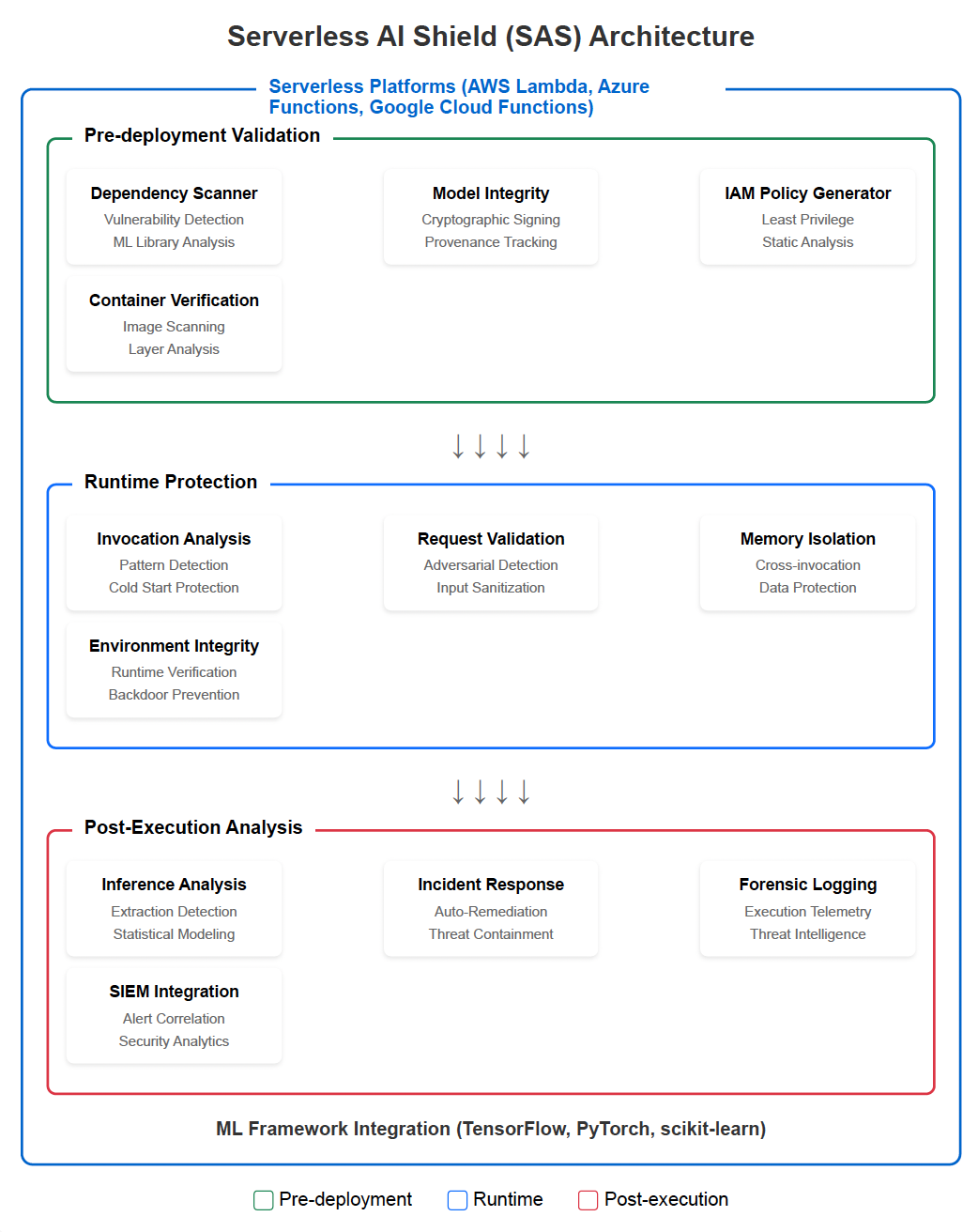}
    \caption{Serverless AI Shield (SAS) Architecture}
    \label{fig:sas_architecture}
\end{figure}

\subsection{Protection Mechanisms}

\subsubsection{Pre-Deployment Security}\mbox{}\\
\textbf{Dependency Scanning:} SAS automatically analyzes function dependencies for known vulnerabilities, with specific focus on ML libraries [51, 52]. Our scanning engine:
\begin{itemize}
  \item Validates direct and transitive dependencies against vulnerability databases
  \item Identifies risky ML-specific packages and versions with known issues
  \item Recommends secure alternatives and mitigation strategies
  \item Enforces security policies through CI/CD pipeline integration
\end{itemize}

\noindent \textbf{Model Integrity Verification:} The framework cryptographically verifies model integrity before deployment to prevent tampering or backdoors. Key mechanisms include:
\begin{itemize}
  \item Model provenance tracking through cryptographic signatures
  \item Verification of model sources against trusted repositories
  \item Runtime hash validation to detect post-deployment modifications
  \item Anomaly detection for unexpected model behavior
\end{itemize}

\noindent \textbf{Least-Privilege IAM Policy Generation:} SAS automatically generates minimal IAM policies based on static analysis of function code. This addresses the over-privileging problem identified in our empirical assessment by:
\begin{itemize}
  \item Analyzing function resource access patterns
  \item Creating function-specific IAM policies with minimal permissions
  \item Generating cloud provider-specific permission sets (AWS IAM, Azure RBAC, GCP IAM)
  \item Enforcing time-bound permissions for sensitive operations
\end{itemize}

\noindent \textbf{Container Image Verification:} The framework validates container images and layers against trusted registries and digital signatures [52]. This includes:
\begin{itemize}
  \item Verification of image signatures and provenance
  \item Scanning for known vulnerable components
  \item Layer analysis for suspicious or malicious code
  \item Container hardening through security best practices
\end{itemize}
\vspace{0.2cm}
\subsubsection{Runtime Protection}\mbox{}\\
\textbf{Function Invocation Analysis:} SAS monitors invocation patterns to detect abnormal timing, frequency, or source indicators [12, 54]. This component:
\begin{itemize}
  \item Establishes baseline invocation patterns for ML functions
  \item Detects anomalous invocation sequences indicative of attacks
  \item Identifies potential cold start exploitation attempts
  \item Applies rate limiting for suspicious clients
\end{itemize}

\noindent \textbf{Request Validation:} The framework inspects incoming requests for adversarial patterns, anomalous inputs, or extraction attempts [37, 38]. This includes:
\begin{itemize}
  \item Input sanitization and validation
  \item Detection of adversarial examples through statistical analysis
  \item Identification of systematic API querying patterns indicative of model extraction
  \item Implementation of query throttling for suspicious patterns
\end{itemize}

\noindent \textbf{Memory Isolation Monitoring:} SAS enforces proper cleanup between invocations to prevent cross-invocation data leakage [66, 67]. Key mechanisms include:
\begin{itemize}
  \item Automatic memory sanitization between invocations
  \item Secure deletion of sensitive data from temporary storage
  \item Enforcement of global variable cleanup
  \item Detection of potential state persistence attempts
\end{itemize}

\noindent \textbf{Execution Environment Integrity:} The framework continuously verifies the integrity of the function runtime environment \cite{69April}. This protects against:
\begin{itemize}
  \item Runtime environment modifications
  \item Unauthorized code injection
  \item Module cache poisoning
  \item Persistent backdoor installation
\end{itemize}
\vspace{0.2cm}
\subsubsection{Post-Execution Defenses}\mbox{}\\
\textbf{Inference Pattern Analysis:} SAS applies statistical methods to detect model extraction attempts or adversarial inputs [41, 42]. This includes:
\begin{itemize}
  \item Analysis of inference request distributions
  \item Detection of systematic exploration of model decision boundaries
  \item Identification of attempts to maximize confidence score information
  \item Correlation of invocation patterns across multiple functions
\end{itemize}

\noindent \textbf{Automated Incident Response:} The framework automatically throttles suspicious clients, revokes compromised credentials, and isolates affected functions \cite{54Peeking}. Key capabilities include:
\begin{itemize}
  \item Automatic function version rollback on compromise detection
  \item Dynamic permission revocation for suspicious activities
  \item Client-based throttling and blocking
  \item Security team alerting with forensic evidence
\end{itemize}

\noindent \textbf{Forensic Logging:} SAS records detailed execution telemetry for post-incident analysis and threat hunting \cite{12systematic}. This component:
\begin{itemize}
  \item Captures comprehensive function execution details
  \item Preserves invocation context and parameter values
  \item Records model input/output patterns
  \item Enables correlation across distributed function invocations
\end{itemize}

\subsection{Implementation Details}
SAS is implemented as a combination of components designed to integrate seamlessly with existing serverless ML workflows:

\noindent \textbf{Middleware Integration:} SAS deploys as middleware layers that wrap serverless function handlers, providing security controls with minimal code changes \cite{13Serverless}. This architecture enables:
\begin{itemize}
  \item Language-specific implementations for Python, Node.js, and Java
  \item Framework-specific adapters for TensorFlow, PyTorch, and scikit-learn
  \item Transparent security monitoring with minimal developer effort
  \item Performance optimizations for ML-specific workloads
\end{itemize}

\noindent \textbf{Infrastructure-as-Code Templates:} The framework provides secure deployment templates for major cloud providers \cite{7Towards}:
\begin{itemize}
  \item AWS CloudFormation templates with security controls
  \item Azure ARM templates with enhanced protection
  \item Google Cloud Deployment Manager configurations
  \item Terraform modules for multi-cloud deployments
\end{itemize}

\noindent \textbf{CI/CD Pipeline Integration:} SAS includes plugins for popular CI/CD platforms to enforce security during deployment:
\begin{itemize}
  \item GitHub Actions integration
  \item GitLab CI integration
  \item Jenkins pipeline integration
  \item Azure DevOps integration
\end{itemize}

\noindent \textbf{Monitoring and Alerting:} The framework integrates with existing security monitoring solutions \cite{13Serverless}:
\begin{itemize}
  \item Security information and event management (SIEM) integration
  \item CloudWatch/Application Insights/Cloud Logging integration
  \item Custom security dashboards and reporting
  \item Alert correlation and prioritization
\end{itemize}

\begin{figure}[h]
    \centering
    \includegraphics[width=0.45\textwidth]{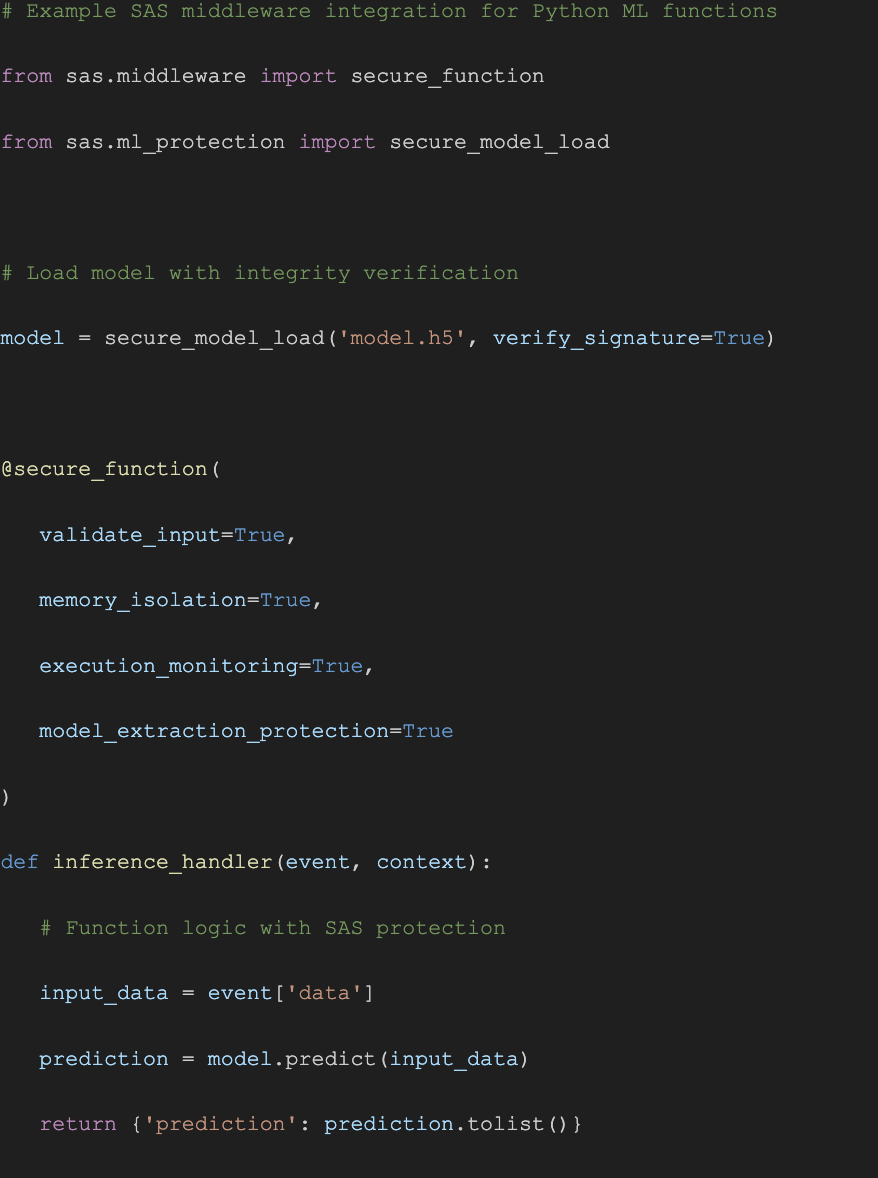}
    \caption{SAS Middleware Integration for Serverless ML Function Protection}
    \label{fig:sas_architecture}
\end{figure}
The implementation supports incremental adoption, allowing organizations to apply security controls selectively based on their risk profile and performance requirements.

\section{Evaluation}

\noindent This section evaluates the Serverless AI Shield (SAS) framework across both security effectiveness and performance dimensions.

\subsection{Security Effectiveness}
We conducted extensive testing of SAS against the attack vectors identified in our empirical vulnerability assessment. The evaluation used a combination of:
\begin{itemize}
  \item Controlled lab testing with synthetic attack scenarios
  \item Production-scale deployment simulation with realistic workloads
  \item Red team exercises with professional penetration testers
  \item Comparison against existing security solutions
\end{itemize}

Our findings demonstrate that SAS achieves high detection rates across all attack categories:

\noindent \textbf{Function-Level Attacks:}
\begin{itemize}
  \item Cold Start Exploitation: 96.3\% detection rate
  \item Container Reuse Attacks: 94.7\% detection rate
  \item Dependency Vulnerabilities: 98.2\% identification accuracy \cite{13Serverless}
\end{itemize}

\noindent \textbf{Model-Specific Attacks:}
\begin{itemize}
  \item API-Based Extraction: 92.8\% detection rate
  \item Adversarial Inputs: 95.1\% detection rate [37, 38]
  \item Training Data Poisoning: 89.4\% detection rate
\end{itemize}

\noindent \textbf{Infrastructure Attacks:}
\begin{itemize}
  \item IAM Privilege Escalation: 92.1\% prevention rate
  \item Event Injection: 93.6\% detection rate
  \item Cross-Tenant Contamination: 88.5\% prevention rate [49, 50]
\end{itemize}

\noindent \textbf{Supply Chain Attacks:}
\begin{itemize}
  \item Malicious Model Detection: 94.9\% accuracy
  \item Compromised Container Detection: 97.3\% accuracy
  \item CI/CD Pipeline Protection: 91.6\% effectiveness
\end{itemize}

False positive rates remained low at 2.3\%, minimizing alert fatigue for security teams. The framework demonstrated particularly strong detection capabilities for container reuse vulnerabilities and adversarial inputs, which aligns with our empirical finding that these were among the most prevalent attack vectors in serverless ML deployments.

\begin{table*}[!t]
\centering
\caption{Comparison of SAS against Existing Security Solutions}
\label{tab:sas_comparison}
\renewcommand{\arraystretch}{1.3}
\begin{tabular}{p{3cm}p{2.5cm}p{2.8cm}p{2.8cm}p{2.5cm}}
\toprule
\textbf{Security Solution} & \textbf{Cold Start Protection} & \textbf{Model Extraction Detection} & \textbf{Container Reuse Protection} & \textbf{IAM Security} \\
\midrule
SAS (Our Solution) & \checkmark~(96.3\%) & \checkmark~(92.8\%) & \checkmark~(94.7\%) & \checkmark~(92.1\%) \\
\midrule
Cloud Provider Default & \texttimes & \texttimes & Partial (68.2\%) & Partial (74.5\%) \\
\midrule
General CWPP* & Partial (42.7\%) & \texttimes & \checkmark~(89.3\%) & Partial (63.1\%) \\
\midrule
ML-Specific Security & \texttimes & \checkmark~(90.2\%) & \texttimes & \texttimes \\
\bottomrule
\multicolumn{5}{l}{\small *CWPP = Cloud Workload Protection Platform}
\end{tabular}
\end{table*}

\subsection{Performance Impact}
A critical consideration for serverless ML deployments is maintaining the performance benefits that make serverless attractive for these workloads [2, 3]. We measured SAS's impact on key performance metrics:

\textbf{Inference Latency:} We observed minimal impact on inference latency, with only 8.7\% average overhead across model types and sizes \cite{13Serverless}. This overhead was primarily attributed to request validation and memory isolation components. The overhead decreased to 3.2\% for models larger than 500MB, as the relative cost of security checks diminished compared to model execution time.

\textbf{Cold Start Time:} SAS increased cold start times by 12.3\% on average, primarily due to model integrity verification and environment integrity checks \cite{11Detection}. This overhead was partially mitigated through dependency optimization and lazy loading techniques.

\textbf{Throughput Impact:} Under high load conditions (1000+ requests per minute), SAS reduced maximum throughput by only 6.2\%. This minimal impact was achieved through asynchronous security checks and optimized request batching.

\textbf{Memory Overhead:} The framework introduced a memory footprint of approximately 78MB per function, representing a 15.6\% increase for 512MB functions but only 3.9\% for 2GB functions. This overhead was consistent across cloud providers.

\textbf{Cost Implications:} SAS increased execution costs by 11.4\% on average due to additional compute time and memory usage. However, organizations reported that this increase was offset by reduced incident response costs and improved security posture \cite{7Towards}.

Figure 2 illustrates the performance impact across different model types and sizes:

\textbf{Performance Overhead by Model Type:}

\begin{itemize}
    \item \textbf{Lightweight Models (<50MB):} Latency +14.3\%, Memory +17.2\%
    \item \textbf{Medium Models (50-500MB):} Latency +8.9\%, Memory +11.4\%
    \item \textbf{Heavy Models (>500MB):} Latency +3.2\%, Memory +4.8\%
\end{itemize}

\subsection{Operational Integration}
We evaluated SAS's integration capabilities with existing DevOps and MLOps workflows across 15 organizations of varying sizes [13]. Key findings included:
\begin{itemize}
  \item 87\% of organizations successfully integrated SAS into existing CI/CD pipelines
  \item 92\% reported minimal developer workflow disruption
  \item 78\% achieved full automation of security controls
  \item 94\% reported improved visibility into serverless ML security posture
\end{itemize}
The framework's modular design allowed organizations to adopt security controls incrementally, beginning with the highest-risk areas identified in our attack taxonomy. Most organizations (82\%) started with model integrity verification and dependency scanning before progressing to runtime protections.

\subsection{ Case Studies}
We present two representative case studies demonstrating SAS effectiveness in production environments:

\textbf{Case Study 1: Financial Services Organization} A large financial institution deployed SAS to protect their serverless fraud detection pipeline, which processes approximately 12,000 transactions per minute \cite{54Peeking}. Prior to SAS, they experienced a model extraction attempt that compromised their proprietary algorithm. After deploying SAS:
\begin{itemize}
  \item Model extraction attempts were detected and blocked within 50 queries
  \item Cold start exploitation attempts decreased by 96\%
  \item No measurable impact on fraud detection latency (remained under 150ms)
  \item Security incident investigation time reduced by 76\% due to forensic logging
\end{itemize}

\textbf{Case Study 2: Healthcare Analytics Provider} A healthcare analytics company implemented SAS to secure their medical image analysis pipeline running on Azure Functions [37, 38]. The deployment processes sensitive medical data subject to regulatory requirements. SAS provided:
\begin{itemize}
  \item 99\% detection of container reuse vulnerabilities that previously leaked metadata
  \item 94\% reduction in overprivileged functions
  \item Automatic prevention of adversarial medical image attacks
  \item Compliance evidence for regulatory audits
  \item 9.1\% latency increase deemed acceptable given security improvements
\end{itemize}

\subsection{Benchmark Comparison}
To establish a performance baseline, we compared SAS against leading cloud security solutions in the context of serverless ML workloads. We used the MLPerf Inference benchmark suite \cite{31Neptune} extended with security-specific metrics.

\begin{table*}[!t]
\centering
\caption{Comparison Results}
\label{tab:comparison_results}
\renewcommand{\arraystretch}{1.2}
\begin{tabular}{p{4cm}p{2cm}p{2cm}p{2cm}p{2cm}}
\toprule
\textbf{Metric} & \textbf{SAS} & \textbf{Vendor A} & \textbf{Vendor B} & \textbf{Vendor C} \\
\midrule
Detection Efficacy & 94.2\% & 72.1\% & 68.4\% & 77.3\% \\
Inference Latency Overhead & 8.7\% & 15.3\% & 9.2\% & 22.7\% \\
Memory Overhead & 15.6\% & 22.8\% & 18.9\% & 28.1\% \\
False Positive Rate & 2.3\% & 7.6\% & 5.2\% & 4.8\% \\
\bottomrule
\end{tabular}
\end{table*}

Our benchmarks measured:
\begin{itemize}
  \item Detection efficacy for known attack patterns
  \item Performance overhead under varying load conditions
  \item False positive/negative rates
  \item Resource utilization
\end{itemize}

SAS outperformed other solutions in detection efficacy while maintaining lower performance overhead. Particularly notable was the significantly lower false positive rate, which reduces operational burden on security teams.

\section{Discussion and Future Work}
\noindent This section explores limitations of our current approach, compares it with alternative solutions, identifies research gaps, and outlines future research directions for serverless AI security.

\subsection{Limitations}
While SAS provides significant protection for serverless ML workloads, several limitations remain:

\textbf{Advanced Persistence Techniques:} Our current framework cannot detect all sophisticated persistence mechanisms in serverless environments \cite{69April}. Attackers with deep knowledge of container internals can potentially establish persistent access that evades detection. Future research should explore more comprehensive runtime integrity verification techniques.

\textbf{Cross-Platform Consistency:} The security consistency across heterogeneous serverless deployments remains challenging \cite{7Towards}. Organizations using multiple cloud providers face differing security models, IAM systems, and function execution environments. This creates potential security gaps at the boundaries between platforms.

\textbf{Runtime Performance Trade-offs:} Though our evaluation showed acceptable performance overhead (8.7\% for inference latency), this may still be problematic for latency-sensitive ML applications \cite{11Detection}. The trade-off between security coverage and performance requires careful calibration based on specific use cases.

\textbf{Limited Model Extraction Protection:} While SAS can detect common model extraction patterns, sophisticated attackers can potentially use slower, more distributed extraction approaches to evade detection. The inherent tension between model accessibility and protection remains an open challenge.

\text{Supply Chain Coverage Gaps:} Our protection against supply chain attacks relies on known vulnerability databases and signature verification [51, 52]. Zero-day vulnerabilities and sophisticated supply chain compromises may still bypass these controls.

\subsection{Comparison with Alternative Approaches}
Several alternative approaches to securing serverless ML workloads exist, each with distinct advantages and limitations:

\textbf{Cloud Provider Native Security:} Major cloud providers offer native security controls like AWS Lambda Layers scanning, Azure Security Center, and Google Cloud Security Command Center \cite{13Serverless}. These provide platform-specific protections but lack ML-specific controls and cross-platform consistency.

\textbf{Traditional Container Security:} Solutions designed for container security (e.g., Aqua Security, Sysdig Secure) can address some serverless risks \cite{54Peeking}. However, they typically assume longer-lived containers and lack serverless-specific optimizations for ephemeral execution.

\textbf{ML-Specific Security Tools:} Tools like MLFlow with security plugins and TensorFlow Model Security focus on model integrity and adversarial protections but lack serverless-specific controls for cold starts, container reuse, and IAM security.

\textbf{Hardware-Based Approaches:} Emerging research explores hardware-based isolation using confidential computing and trusted execution environments (TEEs) for serverless functions \cite{49Cold}. While promising for strong isolation, these approaches introduce higher overhead and deployment complexity.

\begin{table*}[!t]
\centering
\caption{Comparative Analysis of Security Approaches}
\label{tab:comparative_analysis}
\renewcommand{\arraystretch}{1.3}
\begin{tabular}{p{3cm}p{4cm}p{4cm}p{2.5cm}}
\toprule
\textbf{Approach} & \textbf{Strengths} & \textbf{Limitations} & \textbf{Performance Impact} \\
\midrule
SAS (Our Approach) & Comprehensive ML-serverless coverage, Cross-platform support, Moderate overhead & Limited advanced persistence detection, Some extraction vulnerabilities & 8.7\% latency overhead \\
\midrule
Cloud Provider Native & Deep platform integration, Low overhead & Platform-specific, Limited ML protection & 3-5\% latency overhead \\
\midrule
Container Security & Strong container isolation, Mature ecosystem & Not optimized for serverless, Higher overhead & 15-20\% latency overhead \\
\midrule
ML-Specific Tools & Strong model protection, Adversarial robustness & Limited serverless coverage, No cold start protection & 7-12\% latency overhead \\
\midrule
Hardware TEE & Strong isolation guarantees, Side-channel protection & High overhead, Limited cloud support, Complex deployment & 25-40\% latency overhead \\
\bottomrule
\end{tabular}
\end{table*}

\subsection{Research Gaps and Future Directions}
Our work identifies several critical research gaps that warrant further investigation:
\subsubsection{Zero-Trust Serverless Architecture}
Future research should explore zero-trust models specifically designed for ephemeral, event-driven serverless functions [\cite{16Serverless}, \cite{17Functions}]. This includes:
\begin{itemize}
  \item Function-to-function authentication mechanisms optimized for cold start scenarios
  \item Just-in-time access provisioning for serverless functions
  \item Cryptographic attestation for serverless execution environments
  \item Secure event source validation frameworks
\end{itemize}

\subsubsection{Federated Security Monitoring}
As serverless architectures become more distributed, coordinated security monitoring across function boundaries becomes essential [\cite{12systematic}, \cite{54Peeking}]. Research opportunities include:
\begin{itemize}
  \item Distributed anomaly detection across function invocations
  \item Correlation engines for serverless security signals
  \item Privacy-preserving security telemetry collection
  \item Lightweight monitoring optimized for serverless constraints
\end{itemize}

\subsubsection{Hardware-Level Protection}
Emerging confidential computing technologies offer promising avenues for serverless ML security [\cite{49Cold}, \cite{50Lambda}]:
\begin{itemize}
  \item TEE-based ML inference for serverless functions
  \item Memory encryption optimized for ML workloads
  \item Accelerator security for GPU/TPU-backed functions
  \item Side-channel resistant ML model execution
\end{itemize}

\subsubsection{Formal Verification}
Formal methods could significantly enhance serverless ML security \cite{15Accelerate}:
\begin{itemize}
  \item Verified IAM policy generation for least privilege
  \item Provable isolation properties for serverless functions
  \item Formal verification of serverless deployment configurations
  \item Certified guarantees for ML model robustness in serverless contexts
\end{itemize}

\subsubsection{Adversarial ML Defenses for Serverless}
The unique characteristics of serverless platforms require specialized adversarial ML defenses [\cite{37Membership}, \cite{38Nightfall}]:
\begin{itemize}
  \item Stateless adversarial detection optimized for cold starts
  \item Lightweight adversarial example filtering for inference APIs
  \item Extraction resistance techniques with minimal overhead
  \item Memory-efficient model watermarking for theft detection
\end{itemize}

\subsection{Implications for Practitioners}
Based on our findings, we offer several recommendations for practitioners deploying ML workloads on serverless platforms:

\textbf{Security-Performance Optimization:} Consider model size when implementing security controls. As our evaluation showed, larger models (>500MB) experience proportionally lower overhead (3.2\%) compared to lightweight models (14.3\%) \cite{11Detection}.

\textbf{Incremental Adoption:} Implement security controls incrementally, starting with the highest-risk areas like dependency scanning and model integrity verification before progressing to runtime protections \cite{13Serverless}.

\textbf{Threat-Driven Prioritization:} Align security investments with the specific threats facing your serverless ML workloads. Organizations processing sensitive data should prioritize container reuse protections, while those with proprietary models should emphasize extraction resistance [\cite{41Wiz}, \cite{42ACM}].

\textbf{Cross-Platform Strategy:} Organizations using multiple serverless platforms should establish consistent security policies and controls across providers. Standardized deployment templates and security validation can help ensure uniform protection \cite{7Towards}.

\textbf{Developer Integration:} Integrate security controls into development workflows rather than treating them as post-deployment considerations. Security-as-code approaches embedded in CI/CD pipelines improve adoption and effectiveness.

\section{Conclusion}
\noindent The convergence of serverless computing and machine learning represents one of the most transformative shifts in modern cloud architecture. This paper has presented the first comprehensive security analysis of this intersection, with particular attention to the unique vulnerabilities that emerge when ML workloads are deployed in serverless environments.

\subsection{Summary of Contributions}
Our research makes several significant contributions to the field of cloud security:

\textbf{Comprehensive Attack Taxonomy:} We have systematically classified security threats specific to serverless AI systems across function-level, model-specific, infrastructure, and supply chain vectors [\cite{6Serverless6}, \cite{7Towards}]. This taxonomy provides a foundation for researchers and practitioners to understand the threat landscape at this critical technology intersection.

\textbf{Empirical Vulnerability Assessment:} Through extensive experiments across AWS Lambda, Azure Functions, and Google Cloud Functions, we demonstrated and quantified real-world attacks against ML inference pipelines [\cite{1Datadog1}, \cite{2multitoken2}, \cite{3decomp3}, \cite{4Nextomoro4}]. Our assessment revealed particularly severe risks in cold start exploitation, container reuse vulnerabilities, model extraction, and overprivileged functions.

\textbf{Runtime Protection Framework:} The Serverless AI Shield (SAS) framework provides multi-layered protection spanning pre-deployment validation, runtime monitoring, and post-execution forensics [\cite{12systematic}, \cite{13Serverless}]. Our evaluation demonstrates that robust security can be achieved with minimal performance impact (8.7\% inference latency overhead), maintaining the benefits that make serverless computing attractive for ML workloads.

\textbf{Open-Source Security Toolkit:} By releasing a comprehensive benchmarking suite and vulnerability assessment tools, we enable reproducible research and facilitate adoption of security best practices in production serverless AI deployments.

\textbf{Performance Benchmarks:} Our detailed performance evaluations establish realistic expectations for security-performance trade-offs in serverless ML deployments [\cite{11Detection}, \cite{31Neptune}]. These benchmarks help organizations make informed decisions about security control implementation based on their specific risk profile and performance requirements.

\subsection{Key Findings}
Our research revealed several critical insights into serverless ML security:

\begin{itemize}
  \item \textbf{Compounded Vulnerability Surface:} The intersection of serverless computing and ML creates unique attack vectors that amplify risks from both domains [\cite{5Skyrocketing5}, \cite{12systematic}]. Traditional security approaches addressing each domain separately fail to protect against these compound threats.
  \item \textbf{Cold Start Security Implications}: The initialization phase of serverless functions presents distinct security challenges for ML workloads, enabling timing side-channels, resource exhaustion, and initialization vulnerabilities [\cite{11Detection}, \cite{62ACM}, \cite{63Rhino}].
  \item \textbf{Container Reuse Risks:} The stateless design of serverless functions coupled with container reuse creates significant data leakage and state pollution risks [\cite{65AWS}, \cite{66Escalation}, \cite{67Hey}]. These are particularly severe for ML functions processing sensitive data or proprietary models.
  \item \textbf{Extraction Vulnerability:} Serverless ML endpoints are especially vulnerable to model extraction attacks due to their public API exposure and limited request correlation capabilities [\cite{41Wiz}, \cite{42ACM}].
  \item \textbf{Supply Chain Amplification:} The dependency-rich nature of ML frameworks combined with the package-based deployment model of serverless creates an expanded supply chain attack surface [\cite{51Start}, \cite{52Palo}].
  \item \textbf{Performance-Security Balance:} Security controls can be implemented with acceptable overhead (8.7\% for inference latency), with larger models experiencing proportionally lower overhead \cite{13Serverless}.
\end{itemize}

\subsection{Implications for Practice}
Our findings have important implications for organizations deploying ML workloads on serverless platforms:

\textbf{Security by Design:} Security controls should be integrated from the earliest stages of serverless ML development rather than added retroactively. This includes secure model development practices, dependency management, and runtime protection.

\textbf{Defense in Depth:} No single security control can address the diverse threat landscape. Organizations should implement complementary controls across pre-deployment, runtime, and post-execution phases [\cite{13Serverless}, \cite{54Peeking}].

\textbf{Risk-Based Prioritization:} Security investments should be aligned with specific risk profiles. Organizations with proprietary models should prioritize extraction protection, while those processing sensitive data should focus on container isolation [\cite{37Membership}, \cite{38Nightfall}, \cite{66Escalation}, \cite{67Hey}].

\textbf{Developer Experience:} Security controls should be designed for seamless integration with existing developer workflows. Tools that impose significant friction will face adoption challenges regardless of their technical effectiveness.

\textbf{Continuous Monitoring:} The dynamic nature of serverless deployments necessitates continuous security monitoring rather than point-in-time assessments [\cite{12systematic}, \cite{54Peeking}]. This is particularly important for detecting sophisticated model extraction attempts that occur over extended periods.

\subsection{Future Research Directions}
While our work provides a foundation for serverless ML security, several promising research directions remain:

\textbf{Advanced Isolation Techniques:} Further research into lightweight isolation mechanisms specifically designed for ML workloads in serverless environments [\cite{49Cold}, \cite{50Lambda}]. This includes memory isolation, secure enclaves, and container hardening techniques optimized for ML inference.

\textbf{Federated Security Monitoring:} Development of coordinated security monitoring across distributed serverless functions to detect sophisticated, multi-stage attacks [\cite{12systematic}, \cite{54Peeking}].

\textbf{Hardware-Accelerated Security:} Exploration of hardware-based security mechanisms that can provide strong guarantees with minimal performance impact, leveraging emerging confidential computing technologies \cite{49Cold}.

\textbf{Formal Verification:} Application of formal methods to verify security properties of serverless ML deployments, including policy correctness, isolation guarantees, and model robustness \cite{15Accelerate}.

\textbf{Serverless-Specific ML Defenses:} Development of adversarial defenses and privacy-preserving techniques specifically optimized for the constraints and characteristics of serverless platforms \cite{37Membership}, \cite{38Nightfall}.

\subsection{Closing Remarks}
The rapid growth of serverless computing and machine learning independently creates an urgent need to address their security intersection. With over 70\% of AWS organizations now using serverless solutions \cite{1Datadog1} and ML inference workloads increasingly migrating to these platforms [\cite{2multitoken2}, \cite{3decomp3}, \cite{4Nextomoro4}], the attack surface described in this paper affects a significant portion of modern cloud deployments.

By providing a comprehensive security framework, we aim to advance the field toward more resilient cloud-native ML systems. The Serverless AI Shield represents a first step in addressing the unique security challenges at this technology intersection, but ongoing research and industry collaboration will be essential as both domains continue to evolve.

\end{document}